\documentclass[12pt]{article}
\usepackage{amssymb, amsfonts, graphicx, hyperref, psfrag, epsfig, mathrsfs, color, bookmark}
\newcommand{\be}{\begin{equation}}
\newcommand{\bea}{\begin{eqnarray}}
\newcommand{\eea}{\end{eqnarray}}
\newcommand{\ba}{\begin{array}}
\newcommand{\ea}{\end{array}}
\newcommand{\ee}{\end{equation}}

\csname @addtoreset\endcsname{equation}{section}
\begin{document}
\begin{titlepage}

\title{\bf Screening of quark-monopole in $\mathcal{N} = 4$ plasma}
\vskip.3in

\author{\normalsize  Wei-shui Xu$^{1}$ and Ding-fang Zeng$^{2}$ \vspace{12pt}\\
  ${}^1${\it\small Department of Physics, Ningbo University}\\
{\it\small Ningbo 315211, P. R. China}\\
  ${}^2${\it\small College of Applied Science, Beijing University of Technology}\\{\it\small Beijing 100022, P.~R.~China}\\
   {\small E-mail: { \it  xuweishui@nbu.edu.cn, ~ dfzeng@bjut.edu.cn      }} }

\date{}

\maketitle
%\pacs{11.25.Tq, 12.38.Mh}

\voffset -.2in \vskip 2cm \centerline{\bf Abstract} \vskip .4cm
We study a quark-monopole bound system moving in $\mathcal{N}=4$ SYM plasma
with a constant velocity by the AdS/CFT correspondence. The screening length of this system is calculated, and is smaller than that of the quark-antiquark bound state.

\vskip 4.0cm \noindent September 2014
\thispagestyle{empty}

\end{titlepage}

\newpage
\section{Introduction}

The gauge/gravity duality \cite{Maldacena:1997re} is a useful tool to study the
physics of quark gluon plasma (QGP). There are many successful research results
along this line. In \cite{Policastro:2001yc}-\cite{Rebhan:2011vd} etc., the
shear viscosity is calculated by this technique. The jet quenching parameter,
originally defined in the phenomenological study of energy loss of a heavy quark
passing through QGP, can be described and computed nonperturbatively
\cite{Liu:2006ug} in the AdS/CFT context. Another interesting issue related to
energy loss is the drag force experienced by a heavy quark moving in the
$\mathcal{N}=4$ supersymmetric Yang-Mills (SYM) plasma, which was first
calculated in \cite{Herzog:2006gh} for a test string dangling from the boundary
of AdS-Schwarzchild background to the black hole
horizon.

Apart from the remarkable jet quenching phenomenon occurred in hadronization of
a single quark, experimentally one also observed that the production of $J/\psi$
mesons in QGP, when compared to that in proton-proton or proton-nucleus
collisions, is suppressed \cite{Alessandro:2004ap}. Such suppression could be
predicted from phenomenological considerations, since the attractive force
between a quark $q$ and an anti-quark $\bar{q}$ should be screened in a
deconfined QGP, and the screened interaction would not bind that $q\bar{q}$
bound state. In lattice QCD, however, it is difficult to carry out computations for the
screening length $L_s$ of a $q\bar{q}$ pair produced in QGP with a high
velocity. The AdS/CFT proposal \cite{Liu:2006nn} (see also
\cite{Chernicoff:2006hi} ) now provides a calculable way of determining $L_s$
(and the binding energy of the moving $q\bar{q}$ system as well), in
$\mathcal{N}=4$ SYM plasma. This study was generalized to other spacetime
dimensions in the ultra-relativistic limit \cite{Caceres:2006ta}. For more
related references one can see the review \cite{CasalderreySolana:2011us}.

To get a better understanding of the screening effect in SYM plasma, it would be
worthwhile to consider the screening lengths of some bound systems other than
the $q\bar{q}$ system. In the $q\bar{q}$ case one finds
$L_s\propto f(v)(1-v^2)^{1/4}$, where $f(v)$ is a function depending mildly on
the velocity of the plasma wind \cite{Liu:2006nn}. A qualitative explanation of why $L_s$ contains
the factor $(1-v^2)^{1/4}$ is that the screening length should
scale as (energy density)$^{-1/4}$, and the energy density will go like
$(1-v^2)^{-1}$ when the wind velocity gets boosted \cite{Liu:2006nn}. As argued in
\cite{Caceres:2006ta}, this scaling behavior is closely related to the conformal
symmetry of $\mathcal{N}=4$ SYM. Thus, one expects that $(1-v^2)^{1/4}$ is a
kind of "kinetic" factor, which should be seen in any bound systems in the hot
$\mathcal{N}=4$ SYM plasma, and the remaining $v$-dependent factor $f(v)$ should
depends on the dynamical details of the system.

In this paper, we present a concrete test of the above prediction, by studying
screening of a quark-monopole bound system moving with a constant velocity $v$
in a thermal $\mathcal{N}=4$ SYM plasma. At zero temperature a quark of mass
$M_q$ can bind with a monopole of mass $M_m=M_q/g$ to form a dyon, which has the
mass $M=M_q\sqrt{1+1/g^2}$ and is smaller than the total mass $M_q+M_m=M_q(1+1/g)$ of a
free quark and a free monopole (here $g=g_{\rm YM}^2/{4\pi}$ is the string
coupling constant). Such a bound system is not too heavy compared to the quark
mass provided we live in a strong coupling regime ($g\sim 1$). The binding
energy of static dyon at zero temperature and finite temperature was previously studied in \cite{Minahan:1998xb} and \cite{Sfetsos:2007nd}\footnote{The potential of a quark-monopole bound state at finite temperature also is revisited in the Appendix A. } respectively. It
was found that the force between the quark $q$ and the monopole $m$ is indeed
attractive, albeit weaker than the binding force within a $q\bar{q}$ bound state. Of
course one cannot directly see any screening effects in that calculation, since
the temperature was set to be zero there. In this work, we will consider the
$qm$ bound state in a hot plasma wind, try to find its screening length $L_s$ and
compare the result with that derived in the $q\bar{q}$ system.

\section{Quark-monopole in SYM plasma}

We begin with the near horizon geometry $AdS_5\times S^5$ of $N$ coincident D3
branes \bea
ds^2=f^{-\frac{1}{2}}(-hdt^2+d\vec{x}^2)+f^{\frac{1}{2}}h^{-1}dr^2+R^2d\Omega_5^2
\label{metric} \eea where $R$ is the AdS radius determined by $R^4=4\pi g
N\alpha'^2$, $f=\frac{R^4}{r^4}$ and $h=1-\frac{r_0^4}{r^4}$. The horizon of
black hole located at $r=r_0$ and its temperature is $T=r_0/\pi R^2$. According
to AdS/CFT, string theory in this background is dual to $\mathcal{N}=4$ SYM
theory at finite temperature.

\begin{figure}[ht]
\centering{\begin{tabular}{cc}
\includegraphics[scale=0.7]{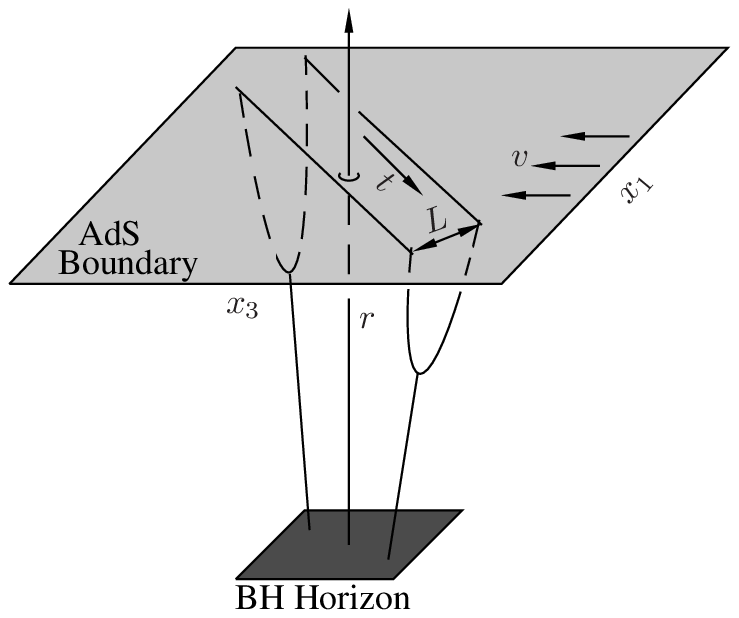} & \includegraphics[scale=0.7]{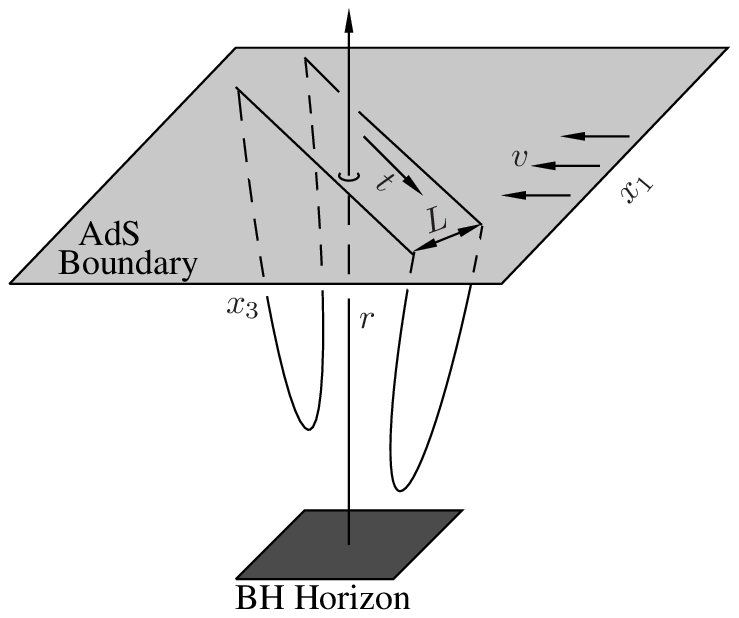}
 \end{tabular}}
\caption{ Left: dual configuration of the $qm$ bound state in a hot plasma wind; the
  dyon is described by a Y-junction of three strings of charged $(1,
   0)$, $(0, 1)$ and $(1, 1)$ with two end points attached on the AdS boundary
  and one end point attach on the black horizon. Right: dual configuration of
   the $q\bar{q}$ system in the same plasma wind; the dipole is described by a
   single fundamental string with both ends attached on the AdS boundary. }
 \label{basicConfiguration}
\end{figure}

Let us consider a dyon moving in the hot $\mathcal{N}=4$ SYM plasma. It is a
bound system of a quark and a monopole, both transforming under the $SU(N)$
fundamental representation. On the gravity side, this system is described by a
fundamental string with charge $(1, 0)$, together with a D-string of
charge $(0, 1)$. Each string has two ends, one of which moves on the AdS
boundary, giving rise to a quark for F-string or a monopole for D-string in the
dual gauge theory, and the other of which lives inside the AdS spacetime. The
ends of F-string and D-string inside the AdS spacetime can be attached to each
other at some junction point to form a bound system. To make the charge
conserved, we have to add a third string of charge $(1, 1)$ to the
system, with one end attached on the junction point of the F- and D-string and
another attached on the horizon of the black hole. The configuration is
therefore described by a Y-junction of three strings with different
charges, as illustrated in the left part of Figure \ref{basicConfiguration}. To be different from the zero temperature case \cite{Minahan:1998xb}, this configuration doesn't preserve supersymmetries at finite temperature.
In order to be the existence of the configuration of Y-junction, the radial coordinate of junction point should be larger than the horizon radius $r_0$. Otherwise, the $(1, 1)$-string in the Y-junction configuration will fall into the horizon of black hole. Then the $(1, 0)$-string and $(0, 1)$-string in the Y-junction configuration will be separated. It means the quark-monopole bound state in the dual gauge theory will be dissolved. This configuration is stable through the stability analysis \cite{Sfetsos:2007nd}.

For comparisons, we shall also consider a $q\bar{q}$ system moving in the same
plasma \cite{Liu:2006nn}, which is simply described by a fundamental string with
both ends attached on the boundary of the AdS spacetime, see the right part of
Figure \ref{basicConfiguration}.

One may choose a frame in which the $qm$ or $q\bar{q}$ bound system is at
rest. This amounts to introduce a plasma wind \cite{Liu:2006nn}. A hot wind in the
$x^3$-direction can be generated by boosting the effective 5-dimensional metric
(\ref{metric}) in the $(t, x^3)$-plane \bea
&&\hspace{-5mm}ds^2=-Adt^2+2Bdtdx^3+Cd{x^3}d{x^3}\nonumber\\
&&\hspace{5mm}+f^{-\frac{1}{2}}(dx^1dx^1+dx^2dx^2)+f^{\frac{1}{2}}h^{-1}dr^2  \label{backgroundMetric} \\
&& A=f^{-\frac{1}{2}}\gamma^2(h-\beta^2),~
B=f^{-\frac{1}{2}}\gamma^2(\beta-\beta h), \cr &&
C=f^{-\frac{1}{2}}\gamma^2(1-\beta^2h), ~~\beta\equiv v, ~\gamma\equiv 1/\sqrt{1-v^2}. \nonumber \eea

We now consider a rest dyon in the velocity-dependent background
(\ref{backgroundMetric}). If the separation between quark and monopole in this dyon is
not along the $x^3$-direction, then the worldsheets of F- and D-string can
be parameterized by \be t=\tau,~~ x^1=\sigma,~~ x^2=\textrm{const.},~~
x^3=x(\sigma),~~ r=r(\sigma). \label{perp} \ee Accordingly, the Nambu-Goto
action for F-string takes the form \be S=-\frac{1}{2\pi\alpha'}\int d\tau
d\sigma\sqrt{-\det g_{\alpha\beta}}=-\frac{\mathcal{T}}{2\pi\alpha'}\int d\sigma
\mathcal{L} \label{actionFstring}\ee where $\mathcal{T}$ is a large time
interval and \be
\mathcal{L}=\sqrt{\gamma^2(h-\beta^2)f^{-1}+hf^{-1}x'^{2}+\gamma^2(1-\frac{\beta^2}{h})r^{\prime2}}.
 \label{lagrangeGeneral}
 \ee The action for D-string can be obtained from (\ref{actionFstring}) by multiplying a factor
 of $1/g=4\pi/g_{\rm YM}^2$. The equation of motion derived from the lagrangian (\ref{lagrangeGeneral})
 can be integrated once with the results
 \be
 x'^2=\frac{p^2}{q^2}\gamma^2(1-\frac{\beta^2}{h}),~ r'^2=\frac{h}{f}
\left[\frac{\gamma^2}{q^2}{(1-\frac{\beta^2}{h})(\frac{h}{f}-p^2)-1}\right]
\label{EqFDstring}
\ee where $p$ and $q$ are integration constants. When $p=0$, we have
$x'(\sigma)=0$ and thus $x^3={\rm const.}$, this particular case describes a
plasma wind blowing perpendicular to the dyon.

If the separation between quark and monopole in the dyon is along the $x^3$ direction,
we may parameterize the F- and D-string as \be t=\tau, ~~ x^{1, 2}={\rm
  const.}, ~~x^3=\sigma,~~ r=r(\sigma). \label{para}\ee Such case corresponds
to the wind blowing parallel to the dyon. With this parameterization, the
lagrangian and the equation of motion read \be
\mathcal{L}=\sqrt{\frac{h}{f}+\gamma^2(1-\frac{\beta^2}{h})r^{\prime2}},
~~~r^{\prime2}=
\frac{h}{\gamma^2(h-\beta^2)}\left[\frac{h^2}{q^2f^2}-\frac{h}{f}\right]
\label{lagrangeParallel}
\ee where $q$ again is an integral constant.

The $(1, 1)$-string is parameterized in a somewhat different way from that of
the F- and D-string. \be t=\tau, ~~~r=\sigma,~~~ x^{1,2,3}=\textrm{const.} \ee
which leads to the following Nambu-Goto action \be
\mathcal{S}=-\frac{\mathcal{T}\sqrt{1+g^{-2}}}{2\pi\alpha'}
\int_{r_0}^{r_j}dr\sqrt{\gamma^2(1-\beta^2h^{-1})}, \ee with $r_0$ is the
horizon of black hole and $r_j$ is the location of the junction point of
strings.

For a $q\bar{q}$ bound state in the background (\ref{backgroundMetric}) the results are
similar. When the dipole is not parallel to the wind direction, the F-string
connecting the quark and anti-quark can be parameterized by (\ref{perp}), so we
get a lagrangian and a set of equations of motion identical to those given in
(\ref{lagrangeGeneral}) and (\ref{EqFDstring}). In the parallel case we can use
the parameterization (\ref{para}) instead, and the corresponding results are
precisely the same as in (\ref{lagrangeParallel}).

Let us consider the plasma wind blowing perpendicular to the dyon (hence
$p=0$). In such case, the first equation in (\ref{EqFDstring}) simply gives
$x(\sigma)={\rm const.}$, while the second reduces to \be
r'^2=\frac{\rho^8}{q^2}\left(\frac{r^4}{r_0^4}-1\right)\left(\frac{r^4}{r_0^4}-\gamma^2-\frac{q^2}{\rho^4}\right),
~~~\rho\equiv\frac{r_0}{R}.\label{rprime}\ee We will write $r'^2=r'^2[r, q]$ to
emphasize the dependence of $r'^2$ on $r$ and $q$. Now the quark and monopole in
this dyon span a distance $L=L_F+L_D$ with \be
L_F\equiv\int_{r_j}^{\infty} dr\frac{1}{\sqrt{r'^2[r,q_F]}},~~~
L_D\equiv \int_{r_j}^\infty dr
\frac{1}{\sqrt{r'^2[r,q_D]}} \label{qFqDandLrelation} \ee where $L_F$ and $L_D$
are the length of F- and D-string projected on the AdS boundary. More
explicitly, one may insert (\ref{rprime}) into (\ref{qFqDandLrelation}) to write
\be L_{F, D}=\frac{r_0q_{F, D}}{\rho^4}\int_{y_j}^\infty
\frac{dy}{\sqrt{(y^4-1)(y^4-\gamma^2-q_{F, D}^2/\rho^4)}}\label{llll}\ee where $ y_j\equiv
\frac{r_j}{r_0}$ and $q_{F, D}\geq 0$. Note that the junction-point
is located at outside the black hole horizon $y_j>1$. Thus, we must choose
$y_j^2\geq \gamma^2+{{\rm max}\{q_F^2, q_D^2\}}/\rho^4$ in order to make both
$L_F$ and $L_D$ be real.

The integrals in (\ref{llll}) can be expressed in terms of the Appell
hypergeometric $F_1$-function. This function, defined through the double
series${}^{}$\footnote{The symbol $(a, n)$ here stands for
 $\Gamma(a+n)/\Gamma(a)$.} \be F_1(a, b, b'; c; \xi,\zeta)=\sum_{m, n=0}^\infty
\frac{(a, m+n)(b, m)(b', n)}{(c,
  m+n)}\frac{\xi^m}{m!}\frac{\zeta^n}{n!},~~~|\xi|<1,
~~|\zeta|<1 \label{ffunction}\ee is the two-variable analogue of the ordinary
Gaussian hypergeometric function $F(a, b; c;\xi)$. In some special cases we will
have $F_1\rightarrow F$. Actually, as $\zeta\rightarrow 0$, only those terms
with $n=0$ will contribute to (\ref{ffunction}), so in this limit $F_1(a, b,
b';c;\xi,0)=F(a,b;c;\xi)$. There exists a simple integral representation for
(\ref{ffunction}) \be F_1(a, b,
b';c;\xi,\zeta)=\frac{\Gamma(c)}{\Gamma(a)\Gamma(c-a)}\int_1^\infty du
u^{b+b'-c}(u-1)^{c-a-1}(u-\xi)^{-b}(u-\zeta)^{-b'}.\label{ffunction1} \ee
clearly, for $b=b'$ this is a symmetric function with respect to $\xi$ and
$\zeta$. Another immediate consequence of (\ref{ffunction1}) is \be F_1(a, b,
b';c;\xi,1)=\frac{\Gamma(c)\Gamma(c-a-b')}{\Gamma(c-a)\Gamma(c-b')}F(a,
b;c-b';\xi). \label{relation}\ee

To find the relation between (\ref{llll}) and (\ref{ffunction1}), we may change
the integration variable $y=y_ju^{1/4}$ in (\ref{llll}) and express $L_{F, D}$
as \be L_{F, D}=\frac{r_0q_{F, D}}{4\rho^4 y_j^3}\int_1^\infty du
u^{-3/4}(u-1)^0\left(u-\frac{1}{y_j^4}\right)^{-1/2}\left(u-\frac{\gamma^2
    +q_{F,D}^2/\rho^4}{y_j^4}\right)^{-1/2}. \ee Comparing this with
(\ref{ffunction1}), we get $a=3/4$, $b=b'=1/2$ and $c=7/4$. One thus obtains
\bea L_F&=&\frac{r_0q_{F}}{3\rho^4
  y_j^3}F_1\left(\frac{3}{4},\frac{1}{2},\frac{1}{2};\frac{7}{4};\frac{1}{y_j^4},
  \frac{\gamma^2+q_F^2/\rho^4}{y_j^4}\right),\cr &&\vspace{10mm} \cr
L_D&=&\frac{r_0q_{D}}{3\rho^4
  y_j^3}F_1\left(\frac{3}{4},\frac{1}{2},\frac{1}{2};\frac{7}{4};\frac{1}{y_j^4},
  \frac{\gamma^2+q_D^2/\rho^4}{y_j^4}\right). \label{lfld}\eea This together with
$L=L_F+L_D$ allows us to determine the distance between the quark and monopole,
in terms of the location $y_j=r_j/r_0$ of the junction point as well as the
integral constants $q_F$ and $q_D$.

Before we proceed to analyze the $qm$ system, let us pause a moment to take a
look at how the Appell function behaves in the $q\bar{q}$ system. If the
plasma wind blows perpendicular to the dipole, the distance $L$ between $q$ and
$\bar{q}$ can be similarly expressed by \bea && L=2\int_{r_j}^\infty dr
\frac{1}{\sqrt{r'^2[r, q]}}\cr
&&\vspace{1.5mm}\cr &&\hspace{5mm}=\frac{2r_0q}{3\rho^4
  y_j^3}F_1\left(\frac{3}{4},\frac{1}{2},\frac{1}{2};\frac{7}{4};\frac{1}{y_j^4},
\frac{\gamma^2+q^2/\rho^4}{y_j^4}\right).
\label{lllll}\eea
One simplicity in the $q\bar{q}$ system is the location of junction point $r_j$
is actually the middle point of a single smooth string. When it passed through
this point along the string, the value of $r'$ changes a sign $r'\rightarrow
-r'$ but does not jump, which implying $r'[r_j, q]=0$. Combining this smoothness
condition with (\ref{rprime}) and the fact that $r_j>r_0$, we see that the
location of the junction point is completely determined, given by
$y_j^4=\gamma^2+q^2/\rho^4$. Thus, the equation (\ref{lllll}) reduces to \be
L=\frac{(2\pi)^{3/2}}{\Gamma(1/4)^2}\frac{r_0}{\rho^2}\frac{q/\rho^2}{(\gamma^2+q^2/\rho^4)^{3/4}}
F\left(\frac{3}{4},\frac{1}{2};\frac{5}{4};\frac{1}{\gamma^2+q^2/\rho^4}\right), \label{llllll}\ee
here we have applied the formula (\ref{relation}). Now for a fixed boost factor
$\gamma$ and considering the asymptotic behavior of $L$ at $q\approx 0$ and
$q\approx \infty$, the result can be directly read off from (\ref{llllll}). For
a small $q$, we have $L\propto q\sim 0$, while for a large $q$, $L\propto
1/\sqrt{q}\sim 0$. So $L$ must have a maximal value $L_{\rm max}$ at some
$q=q_m$, and this gives the screen length $L_s=L_{\rm max}$. To see the velocity
dependence of $L_s$ analytically, we have to take the ultra-relativistic limit
$\gamma\rightarrow \infty$, under which the hypergeometric function in
(\ref{llllll}) behaves as $F=1+\mathcal{O}((\gamma^2+q^2/\rho^4)^{-1})$. So at
the leading order we have $L\propto q(\gamma^2+q^2/\rho^4)^{-3/4}$, which
implying $q_m=\sqrt{2}\gamma\rho^2$ and therefore we get \be
L_s=\frac{r_0}{\rho^2}f_{q\bar{q}}(1-v^2)^{1/4}, ~~~~ f_{q\bar q}\approx
\frac{4\pi^{3/2}}{3^{3/4}\Gamma(1/4)^2}.\label{qqq}\ee The numerical result of
\cite{Liu:2006nn} shows that (\ref{qqq}) holds even beyond the ultra-relativistic limit,
with $f_{q\bar q}=f_{q\bar q}(v)$ being now a function mildly depending on $v$.

Returning to the quark-monopole system, we notice that in general it is not possible to impose the smoothness condition at the Y-junction point $r_j$, and
in particular $r'$ may have a jump when going from F-string to D-string. The
correct condition to determine $y_j$ is that the net force at the string
junction should vanish \cite{Minahan:1998xb} (otherwise the junction point would
move away to lower the energy). Recall that the force exerted by a string at
some point is described by $F^I=\hat{T}E^I_Adx^A/ds$, where $\hat{T}$ denotes
the effective string tension at that point, and $E^I_A$ is a set of vierbeins
associated to the spacetime metric $ds^2=G_{AB}dx^Adx^B$. The tension $\hat{T}$
measures energy per unit length along the string, hence
$\hat{T}ds=(2\pi\alpha')^{-1}\mathcal{L}d\sigma$. We will now evaluate $F^I$ at
the Y-junction point exerted by each string. So we set $T_{(1, 0)}$, $T_{(0, 1)}$
and $T_{(1, 1)}$ to be the tensions of the F-, D- and $(1, 1)$-string,
respectively, at $r=r_j$. For the F-string we have $x_1=\sigma$ and
$r=r(\sigma)$, where $r$ is the solution of (\ref{rprime}) with $q=q_F$. The
infinitesimal length along this string is given by \be ds^2=(f^{-1/2}+f^{1/2}h^{-1}r'^2)d\sigma^2=\frac{\rho^6y_j^2(y_j^4
  -\gamma^2)}{q_F^2}d\sigma^2. \ee On the other hand, the Lagrangian
(\ref{lagrangeGeneral}) with $x_3'=0$ can be evaluated as \bea &&\hspace{-4mm}\mathcal{L}=\gamma
f^{-1/4}(h-\beta^2)^{1/2}(f^{-1/2}+f^{1/2}h^{-1}r'^2)^{1/2}\cr  && =\gamma
f^{-1/4}(h-\beta^2)^{1/2}\frac{ds}{d\sigma},\eea from which one immediately get
\be T_{(1, 0)}=\frac{\gamma
  f^{-1/4}}{2\pi\alpha'}\sqrt{h-\beta^2}=\frac{\rho}{2\pi\alpha'
  y_j}\sqrt{y_j^4-\gamma^2}.\ee Thus, the force $\vec{F}_{(1, 0)}$ exerted by
the F-string at $r=r_j$ has two non-vanishing components, which are determined
by \bea &&\hspace{-10mm} F^1_{(1, 0)}=T_{(1,
  0)}f^{-1/4}\frac{dx^1}{ds}=-\frac{q_F}{2\pi\alpha'\rho y_j},\cr &&
\vspace{6mm} \cr && \hspace{-10mm} F^r_{(1, 0)}=T_{(1,
  0)}f^{1/4}h^{-1/2}\frac{dr}{ds}=\frac{\rho}{2\pi\alpha'
  y_j}\sqrt{y_j^4-\gamma^2-q_F^2/\rho^4}.\eea A similar computation applies to
the D- and $(1, 1)$-string. It is easy to derive, for example, $T_{(0,
  1)}=T_{(1, 0)}/g$, $T_{(1, 1)}=T_{(1, 0)}\sqrt{1+g^{-2}}$. The final result of
$\vec{F}_{(0, 1)}$ and $\vec{F}_{(1, 1)}$ reads \bea && \hspace{-5mm} F^1_{(0,
  1)}=\frac{q_D}{2\pi\alpha'\rho y_j g}, ~~~F^r_{(0, 1)}=\frac{\rho}{2\pi\alpha'
  y_j g}\sqrt{y_j^4-\gamma^2-q_D^2/\rho^4} ,\cr &&\vspace{10mm}\cr && \hspace{-5mm}F^1_{(1,
  1)}=0, ~~~~~F^r_{(1, 1)}=-\frac{\rho\sqrt{1+g^{-2}}}{2\pi\alpha' y_j
  }\sqrt{y_j^4-\gamma^2}. \eea

Having found these forces, we are now ready to impose the condition
$\vec{F}_{(1, 0)}+\vec{F}_{(0, 1)}+\vec{F}_{(1, 1)}=0$. The $x^1$-component of
this condition gives a simple relation between $q_F$ and $q_D$, while the
$r$-component can be used to determine $y_j$ in terms of $q_F$ and
$q_D$. Explicitly, we have \be q_D=gq_F,
~~~y_j^4=\gamma^2+(1+g^2)\frac{q_F^2}{\rho^4}=\gamma^2+\frac{q_F^2+q_D^2}{\rho^4}.\label{relations}\ee
Thus, the expression for $y_j^4$ looks quite similar to that in the $q\bar {q}$
system. It is interesting to note that the location of the junction point does
not change under the S-duality transformation $g\leftrightarrow 1/g$ and
$q_F\leftrightarrow q_D$.

\begin{figure}[ht]
 \centering{\begin{tabular}{cc}
 \includegraphics[scale=0.55]{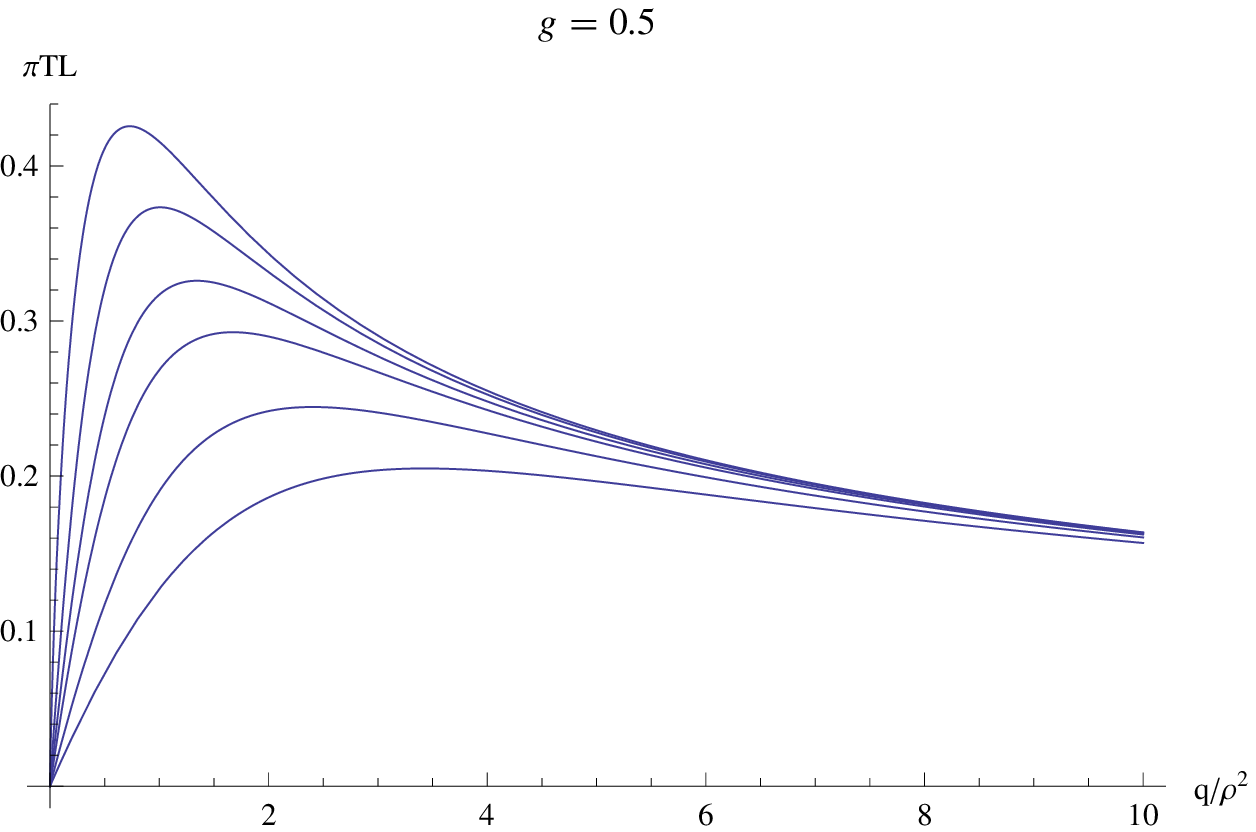} & \includegraphics[scale=0.55]{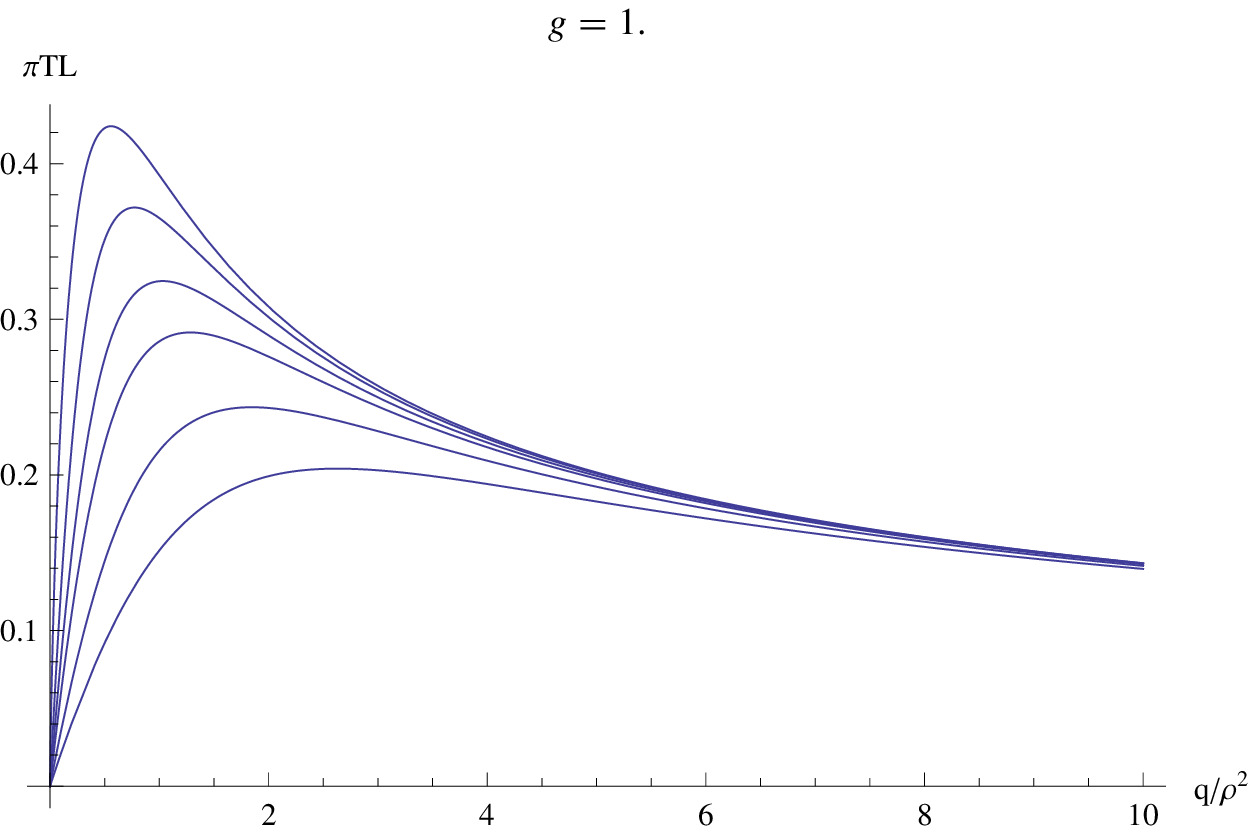}\\
 \includegraphics[scale=0.55]{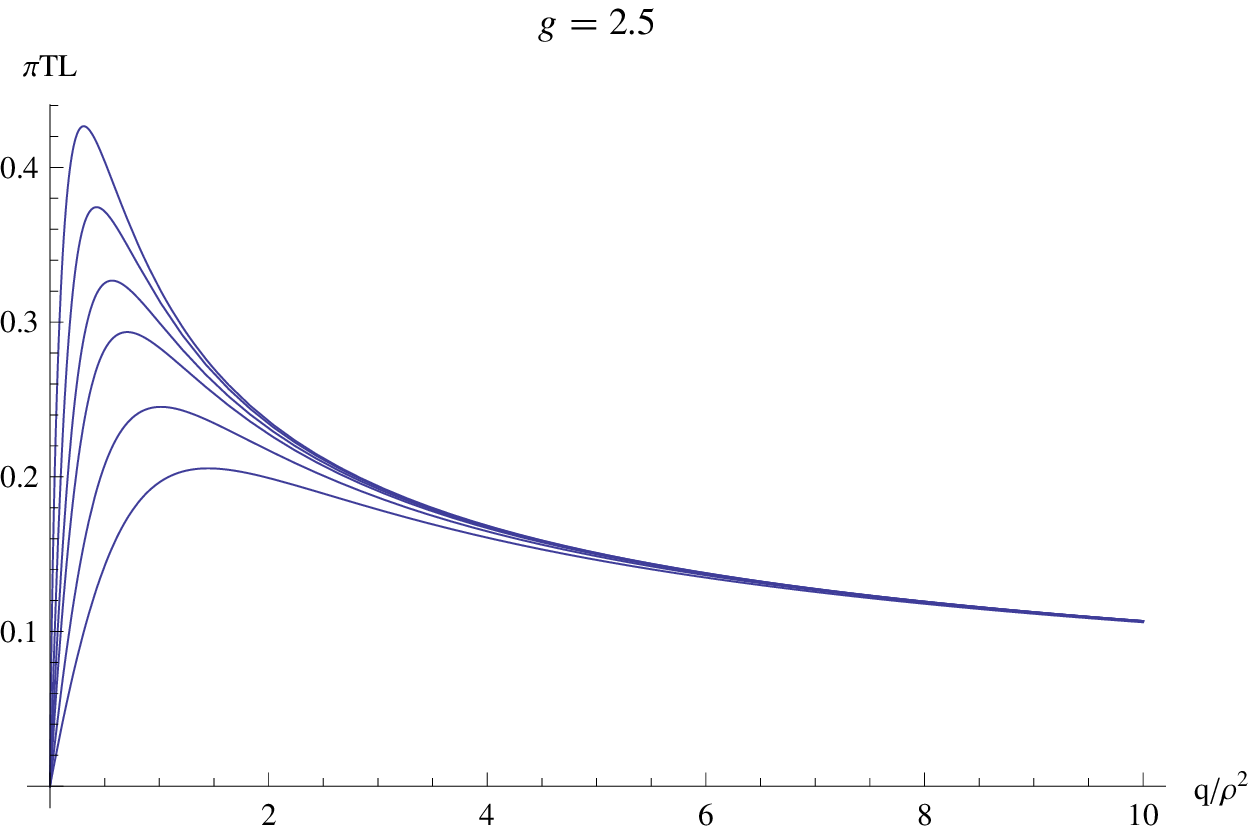} & \includegraphics[scale=0.55]{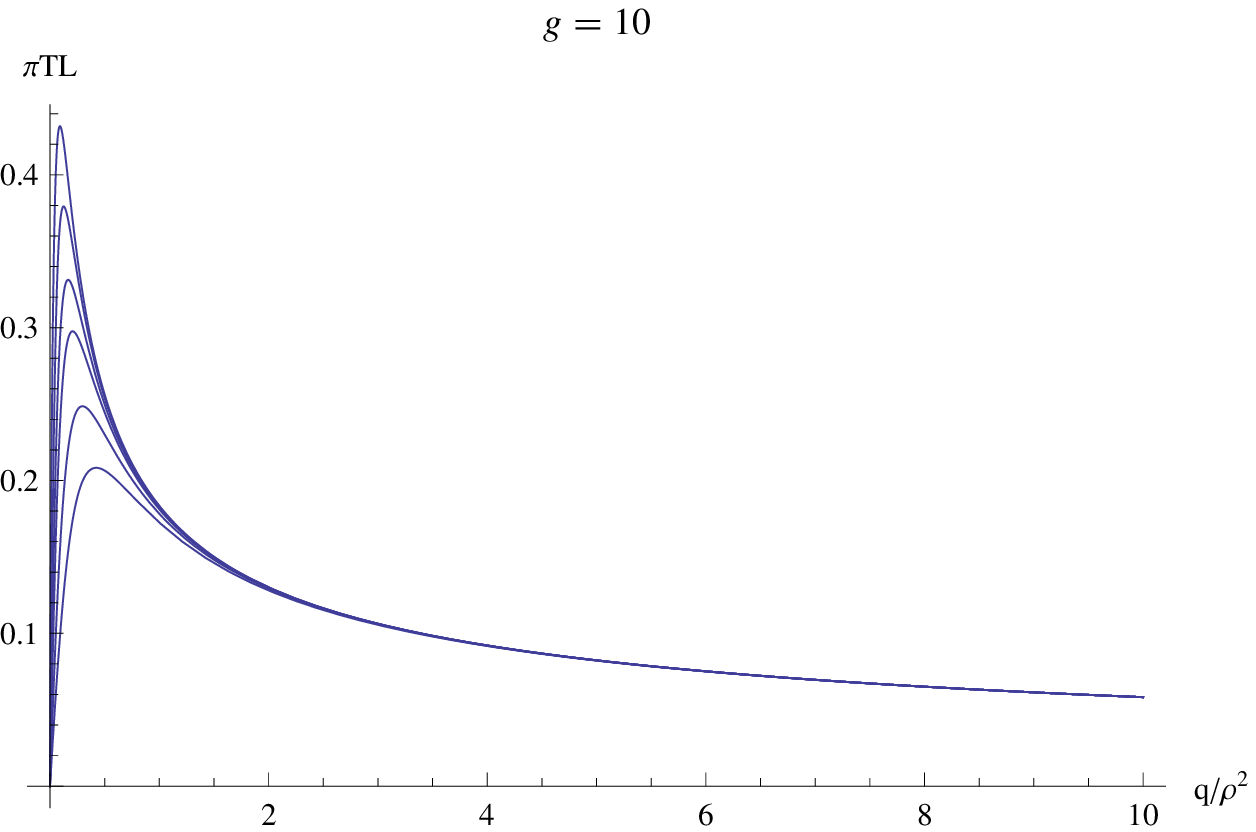}
 \end{tabular}}
 \caption{ Plots of $l\equiv \rho^2L/r_0=\pi TL$ as a function of $q/\rho^2$ at
   $g=0.5,~ 1.0, ~2.5$ and $10$, respectively, for $v=0,~
   0.5,~0.7,~0.8,~0.9,~0.95$ (top to bottom). }
 \label{lqg}
\end{figure}

\begin{figure}[ht]
 \centering{\begin{tabular}{cc}
 \includegraphics[scale=0.55]{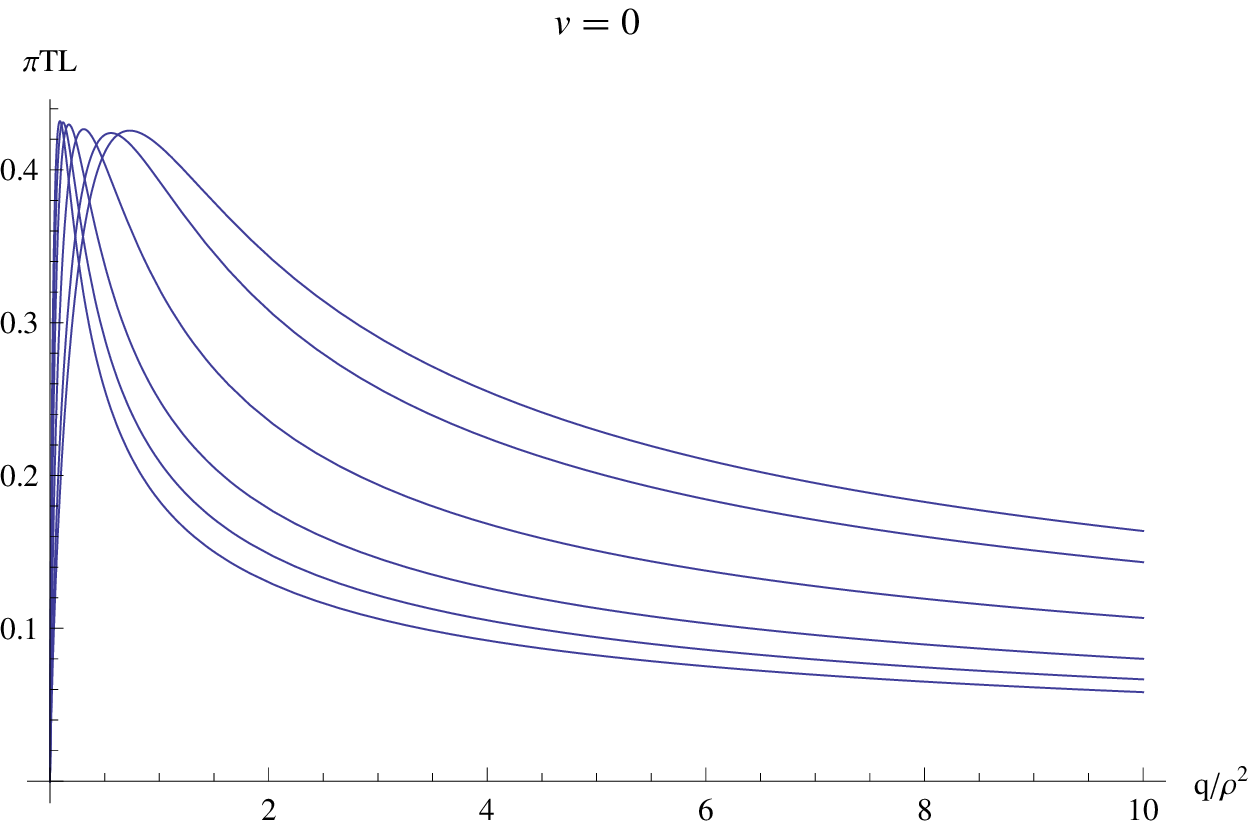}& \includegraphics[scale=0.55]{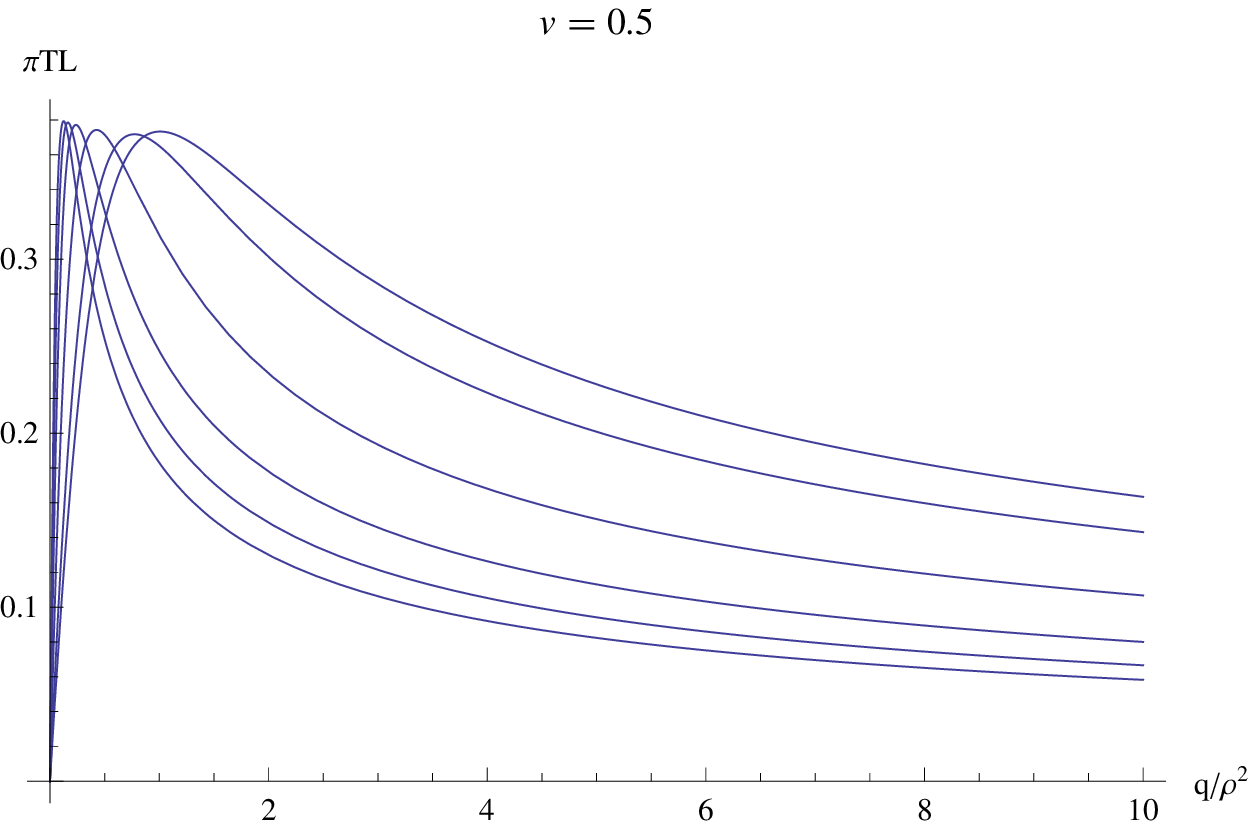}\\
 \includegraphics[scale=0.55]{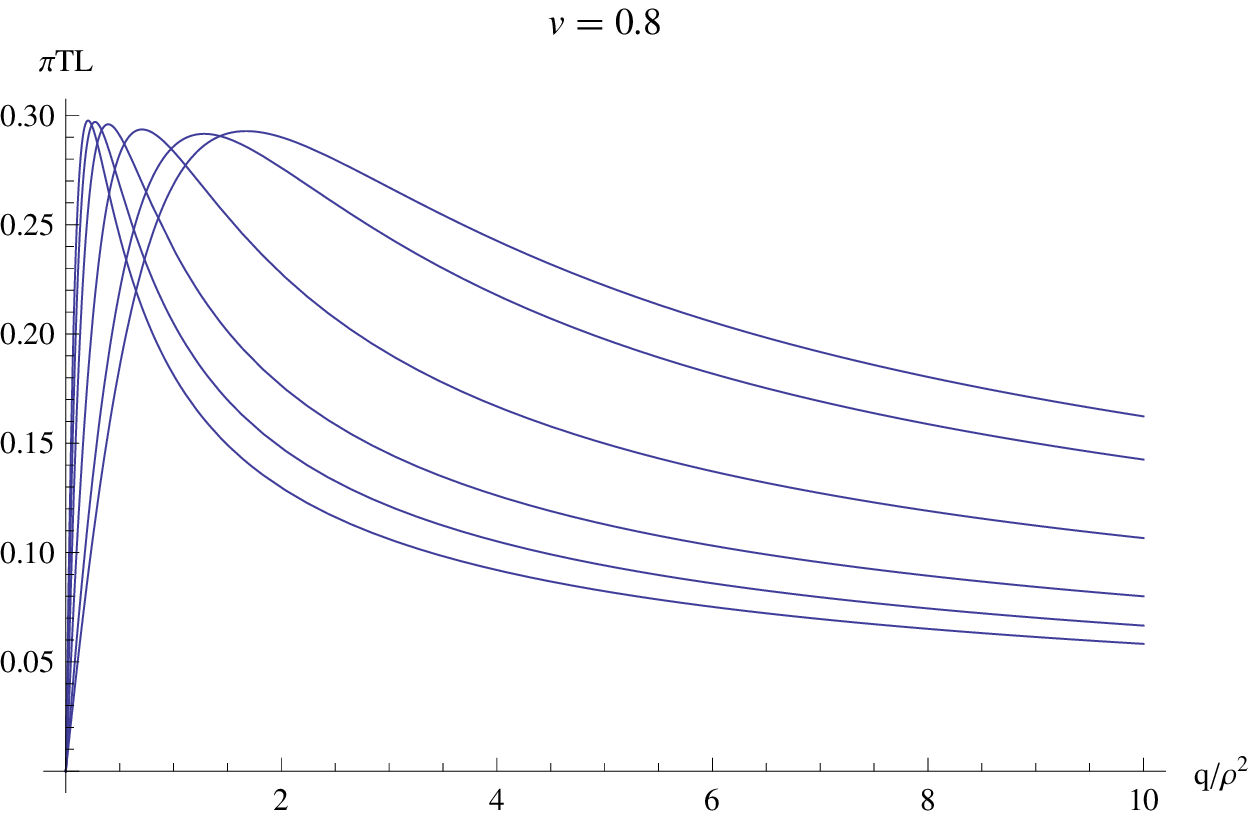}& \includegraphics[scale=0.55]{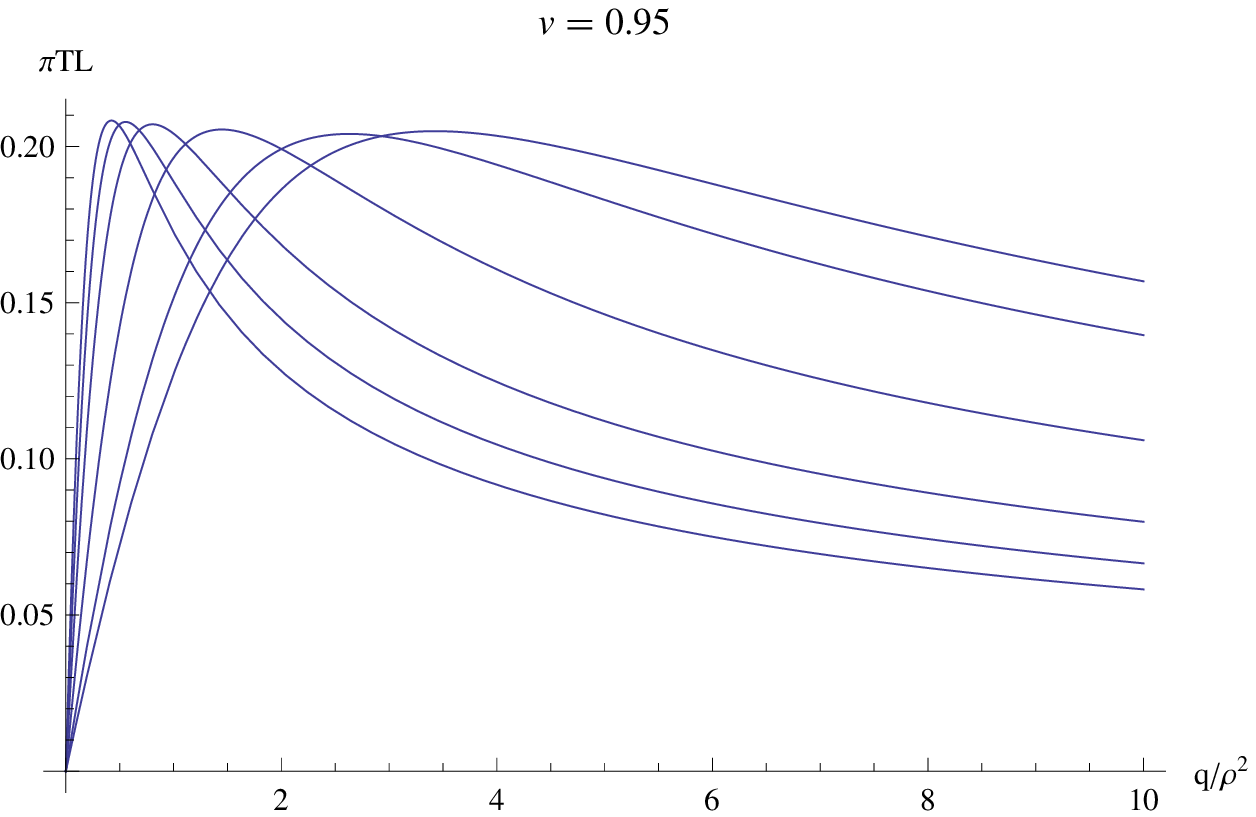}
  \end{tabular}}
  \caption{ Plots of $l\equiv \rho^2L/r_0=\pi TL$ as a function of $q/\rho^2$ at
    $v=0,~ 0.5,~ 0.8$ and $0.95$, respectively, for $g=0.5,~1.0,~2.5,~5.0,~7.5,
    ~10$ (right to left). }
 \label{lqv}
 \end{figure}

\begin{figure}[ht]
 \centering{\begin{tabular}{cc}
 \includegraphics[scale=0.55]{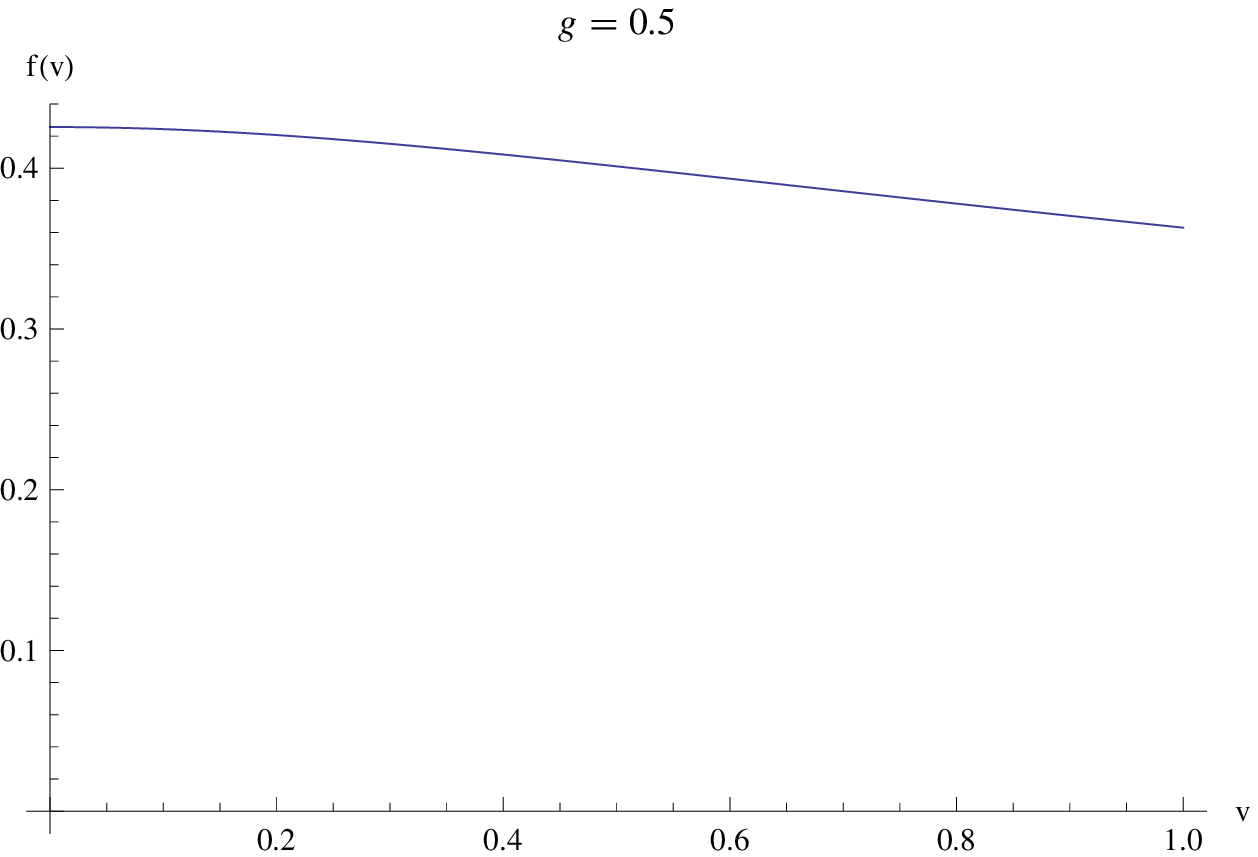}& \includegraphics[scale=0.55]{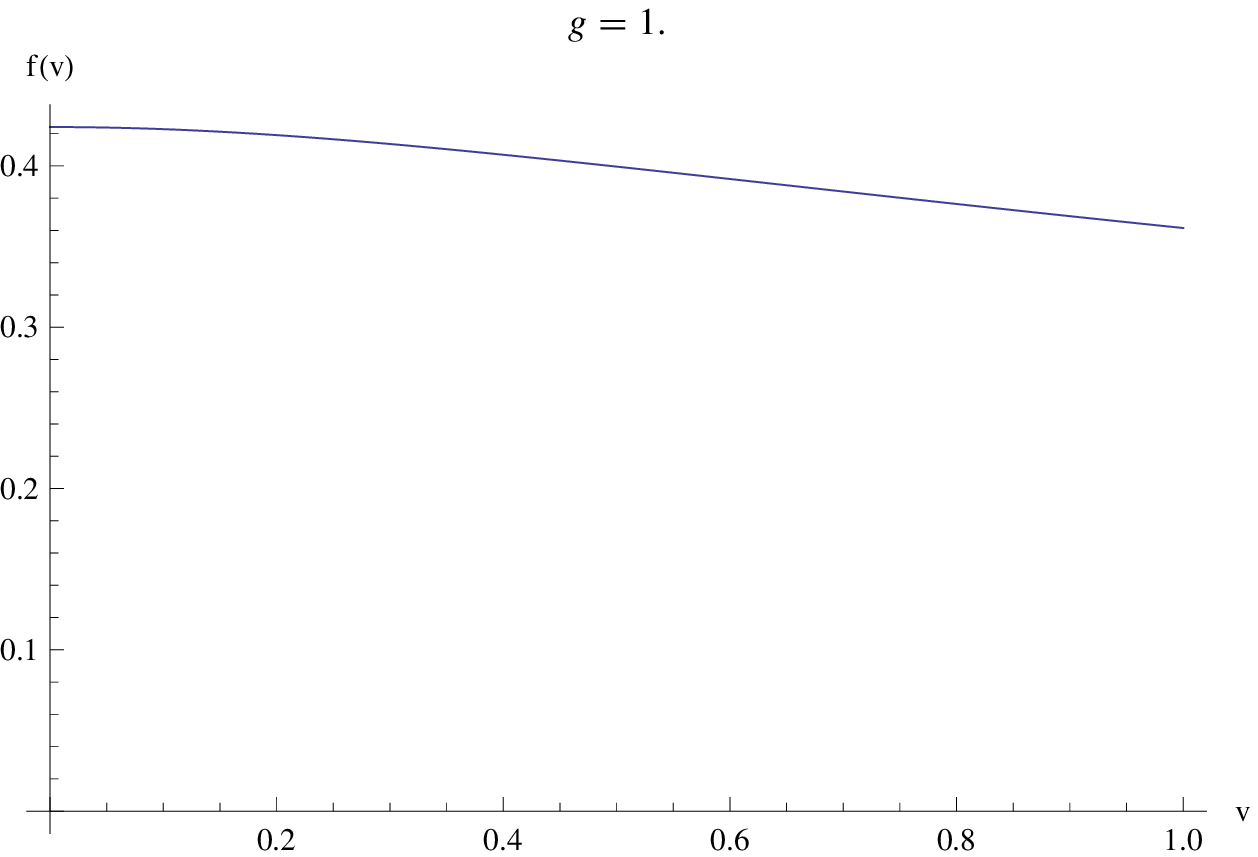}\\
 \includegraphics[scale=0.55]{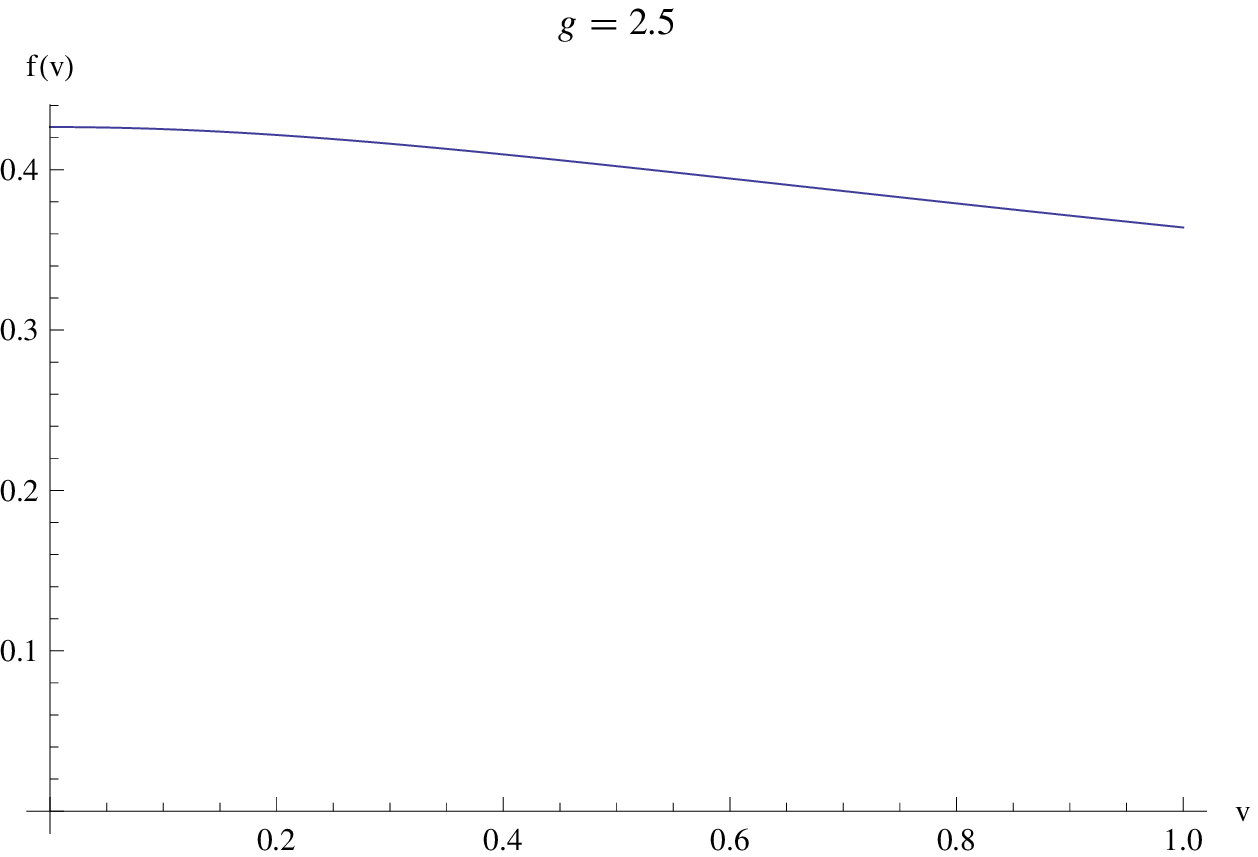}& \includegraphics[scale=0.55]{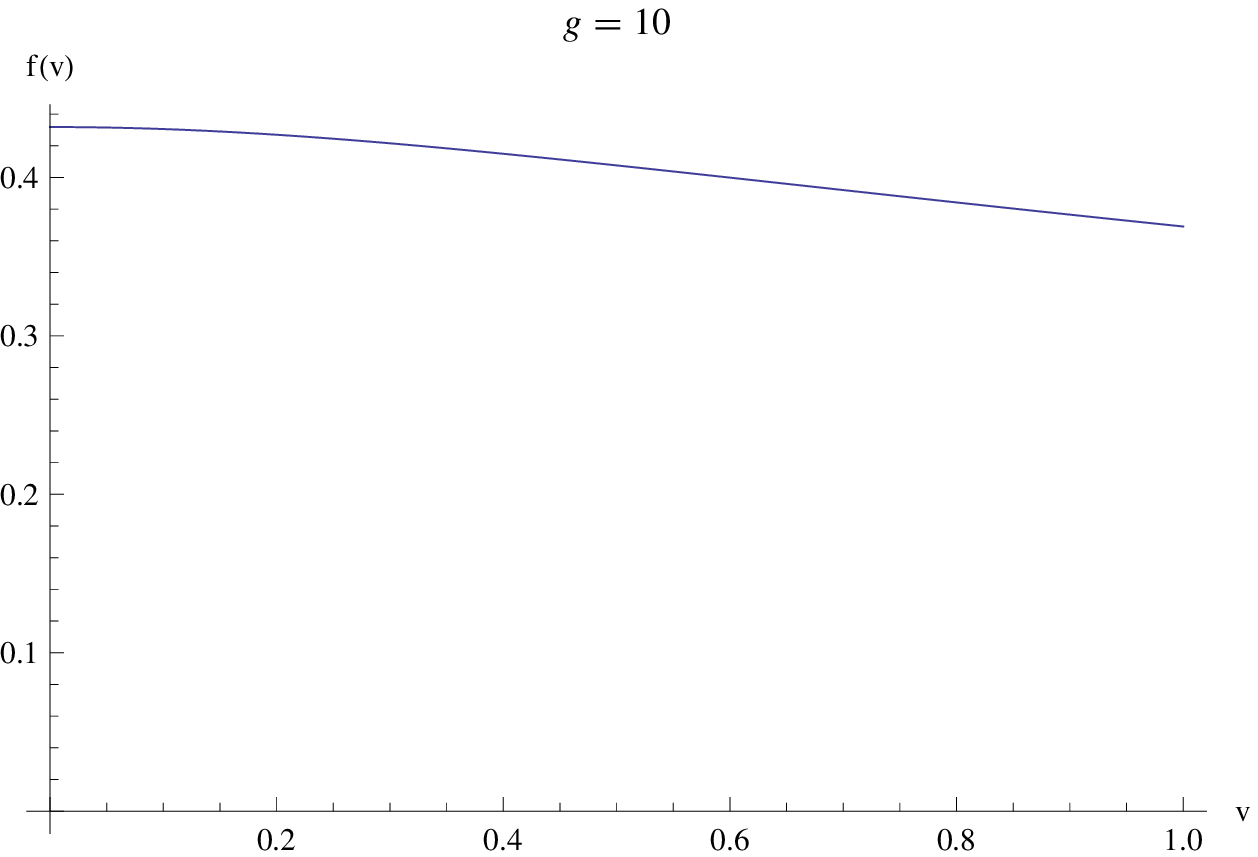}
  \end{tabular}}
 \caption{ The dependence of $f(v,g)$ on $v$ for $g=0.5,~1.0,~2.5, ~10$ is respectively plotted.
 }
 \label{fvg}
\end{figure}

One can use the equation (\ref{relations}) to eliminate the dependence of $L$
on $y_j$ and $q_D$, and express this distance as a single-variable function in
$q_F\equiv q$. The screening effect can be analyzed by looking at the maximal
value of $L=L(q)$ at some $q=q_m$, in analog to the $q\bar{q}$ case
\cite{Liu:2006nn}. After substituting (\ref{relations}) into
(\ref{lfld}), we obtain \bea &&\hspace{-7mm}
L=\frac{r_0}{3\rho^2}\frac{q/\rho^2}{[\gamma^2+(1+g^2)q^2/\rho^4]^{3/4}}\cdot \cr && \vspace{1.5mm}\cr && \left[F_1\left(\frac{3}{4},\frac{1}{2},
    \frac{1}{2};\frac{7}{4};\frac{1}{\gamma^2+(1+g^2)q^2/\rho^4},\frac{\gamma^2+q^2/\rho^4}{\gamma^2+(1+g^2)q^2/\rho^4}
  \right)\right.\label{ll}\\ && \vspace{1.5mm}\cr &&
+\left.gF_1\left(\frac{3}{4},\frac{1}{2},
    \frac{1}{2};\frac{7}{4};\frac{1}{\gamma^2+(1+g^2)q^2/\rho^4},
    \frac{\gamma^2+g^2q^2/\rho^4}{\gamma^2+(1+g^2)q^2/\rho^4}\right)\right].\nonumber\eea
One may fix the boost factor $\gamma$ and examine the asymptotic behavior of $L$
in the small and large $q$ regions, as in the $q\bar{q}$ case. When
$q\rightarrow 0$, the two $F_1$ functions in (\ref{ll}) behave smoothly, both
approaching to the $\gamma$-dependent constant \be
F_1\left(\frac{3}{4},\frac{1}{2},
  \frac{1}{2};\frac{7}{4};\frac{1}{\gamma^2},1\right)=\frac{3(2\pi)^{3/2}}{2\Gamma(1/4)^2}F
\left(\frac{3}{4},\frac{1}{2}; \frac{5}{4};\frac{1}{\gamma^2}\right)\ee where we have used the formula (\ref{relation}). So we
find $L\propto q\rightarrow 0$ in this limit. Similarly we see that in the limit
$q\rightarrow \infty$, then $L\propto 1/\sqrt{q}\rightarrow 0$. Thus, $L=L(q)$
is a function positive everywhere, it must have a maximal value $L_{\rm max}$ at
some extremal point $q=q_m$. For convenience, we define a dimensionless quantity $\pi T L$. Then, through some numerical calculations, we show $\pi T L$ (at fixed temperature) to depend on the parameter $q/\rho^2$ at fixed coupling constant $g$ and velocity $v$ in Fig. \ref{lqg} and Fig. \ref{lqv}.
 These two figures indicate that the quark-monopole system indeed has a screening length
 $L_s=L_{max}$. In addition, we find that
 (i) the screening length $L_s$ of the $qm$ system is smaller than that of the
 $q\bar{q}$ pair, and (ii) in the $qm$ case, the dependence of $L_s$ on the
 coupling constant $g$ is rather mild. In order to show the dependence of $L_s$ on $(1-v^2)^{1/4}$, we define
$f(v,g)\equiv (1-v^2)^{-1/4}\pi TL_s$. Then, the dependence of $f(v, g)$ on the parameters $v$ and $g$ is plotted in Fig. \ref{fvg}.
 It shows that this dependence on the parameter $g$ is mild, and its dependence on $v$ is similar to the $q\bar{q}$ case. This provides an explicit test of the prediction mentioned in the
introduction: $(1-v^2)^{1/4}$ is a kind of "kinetic" factor that can be seen in
any bound systems in the $\mathcal{N}=4$ hot plasma.

It is possible to derive the ultra-relativistic behavior of the screening length
analytically. Let us take the large $\gamma$ limit and approximate the equation (\ref{ll}) by
\bea &&\hspace{-10mm}
L=\frac{r_0}{3\rho^2}\frac{q/\rho^2}{[\gamma^2+(1+g^2)q^2/\rho^4]^{3/4}}\cdot \cr &&\vspace{1.5mm}\cr &&\left[F\left(\frac{3}{4},
    \frac{1}{2};\frac{7}{4};\frac{\gamma^2+q^2/\rho^4}{\gamma^2+(1+g^2)q^2/\rho^4}
  \right)\right.\label{lll}\\ && \vspace{1.5mm}\cr &&
+\left.gF\left(\frac{3}{4},
    \frac{1}{2};\frac{7}{4};\frac{\gamma^2+g^2q^2/\rho^4}{\gamma^2+(1+g^2)q^2/\rho^4}\right)\right].\nonumber\eea
One may consider a range of $q$ behaves as $q\sim \gamma^\alpha u$ with some
fixed number $\alpha$ and a rescaled variable $u\sim \mathcal{O}(\gamma^0)$. It
is not difficult to see that such a range does not contain the extremal point
$q_m$ of $L$, unless $\alpha=1$. In fact, if $\alpha\neq 1$, each hypergeometric
function in (\ref{lll}) will tend to a constant be independent of $u$ in the
limit $\gamma\rightarrow \infty$, so that $L$ can be further approximated by \bea &&
L(q)\propto \frac{q/\rho^2}{[\gamma^2+(1+g^2)q^2/\rho^4]^{3/4}}\cr&& \vspace{3mm} \cr  &&\hspace{5mm}\Longrightarrow
L'(q)\propto
\frac{2\gamma^2-(1+g^2)q^2/\rho^4}{[\gamma^2+(1+g^2)q^2/\rho^4]^{7/4}}.\eea It
follows that $L'(q)$ never vanishes in that range. Thus, the extremal point
$q_m$ has to scale as $q_m=\gamma u_m$ with $u_m\sim
{\mathcal{O}}(\gamma^0)$. Substituting this into (\ref{lll}) we obtain the
scaling behavior of the screening length $L_s=L(q_m)\sim (1-v^2)^{1/4}$ in the
large $\gamma$ regime.

\section{Summaries}

We consider a quark-monopole system through using its gravity dual description. In the gravity side, this configuration includes F-string, D-string and $(1, 1)$-string, which are connected at a junction point. We calculate the screening length of quark-monopole bound state moving in a hot $\mathcal{N}=4$ SYM
plasma. We find the screening length $L_s$ is smaller than that of the quark-antiquark bound state. And its dominant dependence of $L_s$ on the wind velocity $v$ is proportional to $(1-v^2)^{1/4}$. Finally, the dependence of screening length $L_s$ on the string coupling constant $g$ is very mild. Thus, it is not very easy to distinguish the quark-antiquark pair from the quark-monopole bound state through calculating the screening length in a hot plasma.

\subsection*{Acknowledgments}

We are very glad to thank Prof. Yi-hong Gao for the
collaboration in the early stage of this project. The work of W.-s. Xu is partly
supported by K. C. Wong Magna Fund in Ningbo University, National Science Foundation of China under Grant No. 11205093 and 11347020. The work of D.-f. Zeng is surpported by BJNSF under Grant No. 1102007.

\appendix

\section{Quark-monopole potential}

In this appendix, we should investigate the quark-monopole potential in the $AdS_5\times S^5$ black hole background (\ref{metric}). Similar computation of this binding potential also is performed in \cite{Sfetsos:2007nd}. We assume the worldsheets of F- and D-string are parameterized by $\tau=t$ and $\sigma= x^1$, then the action for F-string is
\be S=\frac{1}{2\pi\alpha'}\int d\tau d\sigma \sqrt{r'^2+ \frac{h}{f}}, \label{aeom}\ee
which can be derived from the equation (\ref{lagrangeGeneral}) by setting the velocity of plasma wind $v=0$.
The action for D-string is got by multiplying the factor $1/g$ on the action of F-string. Then the equation of motion reads \be  r'^2=\frac{h^2}{q_{F, D}^2f^2}-\frac{h}{f}\ee with the integral constants $q_F$ and $q_D$ for F- and D-string respectively. From the equation (\ref{llll}),  the lengths of F- and D-string are  \be L_{F, D}=\int_{r_j}^\infty \frac{dr}{\sqrt{\frac{h^2}{q_{F, D}^2f^2}-\frac{h}{f}}}=\frac{r_0q_{F, D}}{\rho^4}\int_{y_j}^\infty
\frac{dy}{\sqrt{(y^4-1)(y^4-1-q_{F, D}^2/\rho^4)}}, \ee where $y_j=r_j/r_0$, and $r_j$ is the junction point of F-, D- and $(1, 1)$-string. Thus, the distance between quark and monopole in the dyon is \bea &&\hspace{-7mm}
L=\frac{r_0}{3\rho^2}\frac{q_F/\rho^2}{[1+(q_F^2+q_D^2)/\rho^4]^{3/4}}\cdot \cr && \vspace{1.5mm}\cr && \left[F_1\left(\frac{3}{4},\frac{1}{2},
    \frac{1}{2};\frac{7}{4};\frac{1}{1+(q_F^2+q_D^2)/\rho^4},\frac{1+q_F^2/\rho^4}{1+(q_F^2+q_D^2)/\rho^4}
  \right)\right.\label{app4}\\ && \vspace{1.5mm}\cr &&
+\left.gF_1\left(\frac{3}{4},\frac{1}{2},
    \frac{1}{2};\frac{7}{4};\frac{1}{1+(q_F^2+q_D^2)/\rho^4},
    \frac{1+q_D^2/\rho^4}{1+(q_F^2+q_D^2)q^2/\rho^4}\right)\right].\nonumber\eea By using the equation (\ref{aeom}) and subtracting the divergence, the potential of quark-monopole is expressed as \bea &&\hspace{-10mm}E_{QM}=\frac{r_0}{2\pi\alpha'}\left[\int_{y_j}^\infty dy\left(\frac{1}{\sqrt{1-q_F^2\frac{f}{h}}}-1\right)-(y_j-1)\right.\cr && \left.+\frac{1}{g}\int_{y_j}^\infty dy\left(\frac{1}{\sqrt{1-q_D^2\frac{f}{h}}}-1\right)-(y_j-1)/g +\sqrt{1+g^{-2}}~(y_j-1)\right].\label{app5}\eea
If $r_0=0$, then the distance $L$ and potential $E_{QM}$ will reduce to the corresponding cases \cite{Minahan:1998xb}. By using the equations (\ref{app4}) and (\ref{app5}), and the vanishing condition of net force \be q_D=gq_F, ~~ y_j^4=1+\frac{q_F^2+q_D^2}{\rho^4}\ee  at junction point of F-, D- and (1, 1)-string, the quark-monopole potential at finite temperature reads
\bea &&\hspace{-14mm}E_{QM}=\frac{\sqrt{4\pi N}}{6\pi L}\,\sqrt{g}\, \frac{q_F/\rho^2}{[1+(q_F^2+q_D^2)/\rho^4]^{3/4}}\cdot \cr && \vspace{1.5mm}\cr && \left[F_1\left(\frac{3}{4},\frac{1}{2},
    \frac{1}{2};\frac{7}{4};\frac{1}{1+(q_F^2+q_D^2)/\rho^4},\frac{1+q_F^2/\rho^4}{1+(q_F^2+q_D^2)/\rho^4}
  \right)\right.\cr &&\vspace{1.5mm}\cr && \hspace{4mm}
+\left. g F_1\left(\frac{3}{4},\frac{1}{2},
    \frac{1}{2};\frac{7}{4};\frac{1}{1+(q_F^2+q_D^2)/\rho^4},
    \frac{1+q_D^2/\rho^4}{1+(q_F^2+q_D^2)/\rho^4}\right)\right]\cdot\cr &&
\left[\int_{y_j}^\infty dy\left(\frac{1}{\sqrt{1-q_F^2\frac{f}{h}}}-1\right)-(y_j-1)\right.\cr && \left. \hspace{3mm}+\frac{1}{g}\int_{y_j}^\infty dy\left(\frac{1}{\sqrt{1-q_D^2\frac{f}{h}}}-1\right)-(y_j-1)/g +\sqrt{1+g^{-2}}~(y_j-1)\right].
\eea
This potential is negative for all coupling constant $g$, which is shown by the left figure of Fig. 5. We also plot the dependence of this binding energy on the temperature in the right figure of Fig. 5. \begin{figure}[ht]\centering{\begin{tabular}{cc}
\includegraphics[scale=0.56]{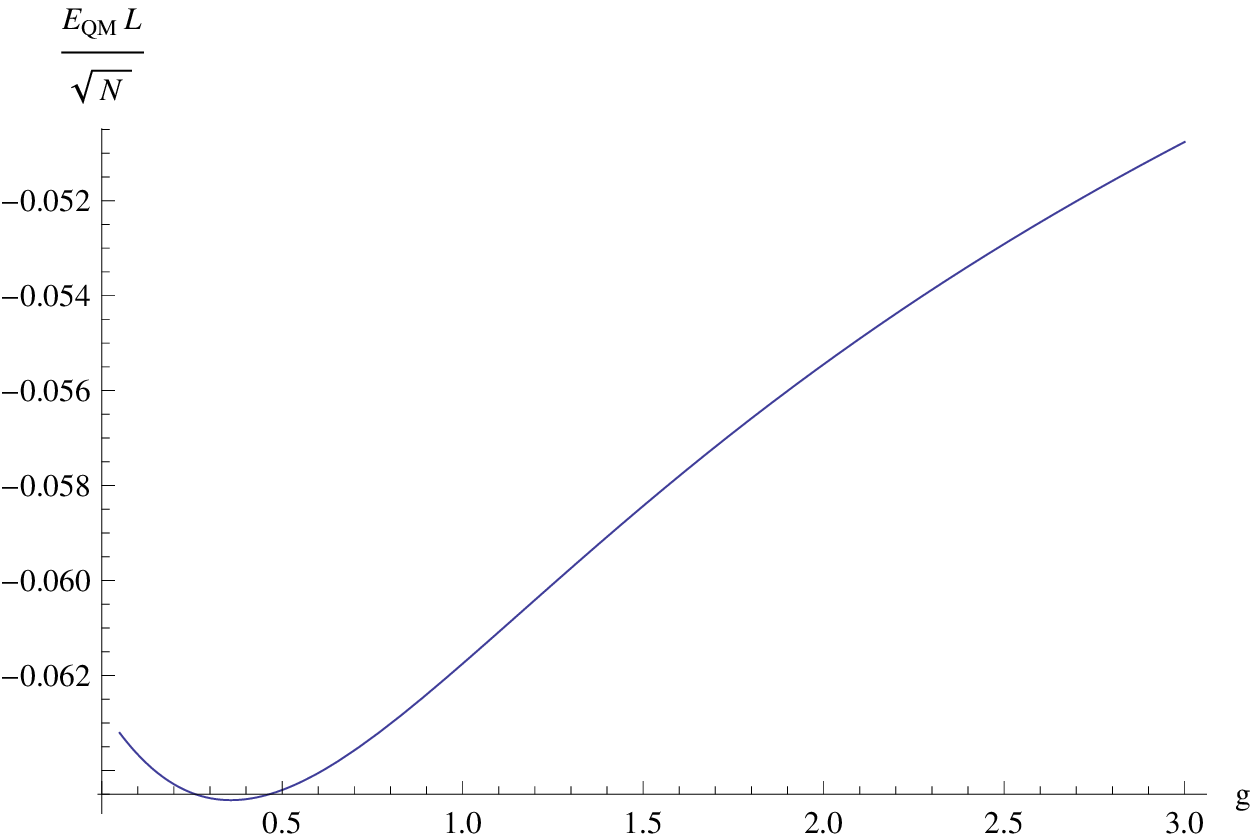} & \includegraphics[scale=0.56]{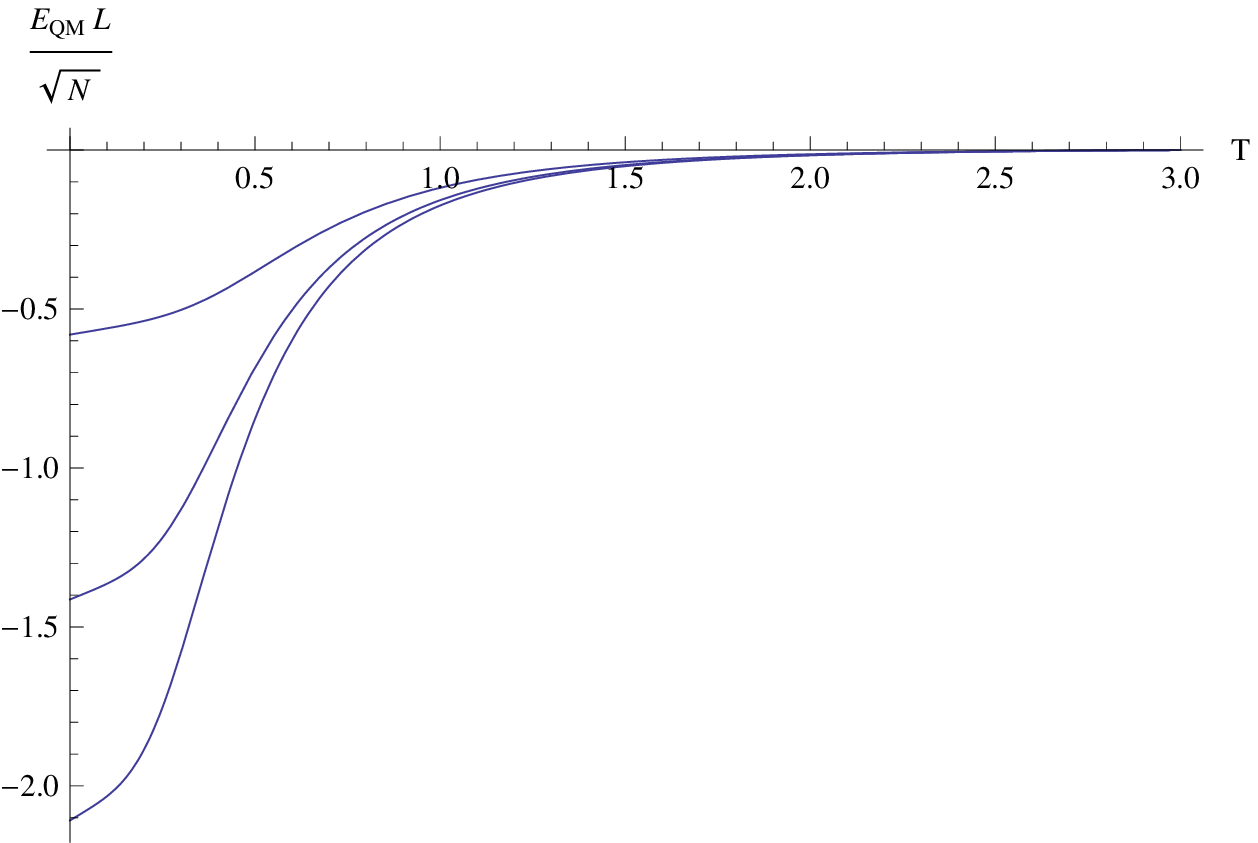}\end{tabular}
 \caption{ The dependence of $\frac{E_{QM}L}{\sqrt{N}}$ on $g$ is plotted with $R=1$, $q_F=1$ and $\rho=10$ on the left.  For the right figure, it shows $\frac{E_{QM}L}{\sqrt{N}}$ depends on the temperature $T$ at $g=0.5,~ 1,~ 2.5$ from bottom to up with setting $R=1$ and $q_F=1$.}}
  \label{ee}
\end{figure}
As expected, the binding energy of quark-monopole will approach to zero as the junction point $r_j$ goes to the horizon of black hole. The reason is now the junction point will pass through the horizon, and the F- and D-sting will be not connected. From the equation (\ref{relations}), we know the junction point $y_j$ is invariant under the S-duality transformation $g\leftrightarrow 1/g$ and $q_F\leftrightarrow q_D$. Thus, the quark-monopole potential at finite temperature is still invariant under the S-duality. Similar to the cases of $q\bar{q}$ and $qm$ at zero temperature, the potential is still proportional to $1/L$ even if the conformal symmetry is broken by the temperature of black hole.

\end{document}